\documentclass[prl,twocolumn,floatfix,reprint,showpacs]{revtex4}
\usepackage{graphicx}
\usepackage{amsmath}
\begin{document}

\title{Observation of Dynamical Super Efimovian Expansion in a Unitary Fermi Gas}

\author{Shujin Deng$^1$, Pengpeng Diao$^1$, Fang Li$^1$ Qianli Yu$^1$, Shi Yu$^1$, and Haibin Wu$^{1*}$}

\affiliation{$^1$State Key Laboratory of Precision Spectroscopy, East China Normal University, Shanghai 200062, P. R. China}


\date{\today}

\begin{abstract}
We report an observation of a dynamical super Efimovian expansion in a two-component strongly interacting Fermi gas by engineering time dependent external harmonic trap frequencies. When trap frequency is followed as $[1/4t^2+1/t^2\lambda\log(t/t_*)]^{1/2}$, where $t_*$ and $\lambda$ are two control parameters, and the change is faster  than  a critical value, the expansion of such the quantum gas shows a novel dynamics due to its spatial and dynamical scaling symmetry.  A clear double-log periodicity, which is a hallmark of the super Efimov effect, is emergent for the cloud size in the expansion. The universality of such scaling dynamics is verified both in the non-interacting limit and in the unitarity limit.  Observing super-Efmovian evolution represents a paradigm in probing universal properties and allows in a new way to study many-body nonequilibrium dynamics with experiments.
\end{abstract}

\maketitle

The understanding of nonequilibrium dynamics in the strongly coupled quantum gases is an open problem at the frontiers of many-body physics with widespread applications. The presence of strong interactions between constituent particles challenges any attempt aimed at controlling their dynamics. Therefore to explore the consequences and implications of symmetrises in such systems becomes very important because any symmetry or invariance can lead to a significant reduction of the complexity of a problem, possibly rendering it solvable.  One example of great interest is the scale invariance which has been widely studied the thermodynamics~\cite{ScaleThermal1,ScaleThermal2} and dynamics~\cite{critialpoint} near critical points, and biological complexes~\cite{Scalebio}. Although scale invariance has had made a success in studying the energetics and critical dynamics of these rather difficult systems, the role of scale invariance on nonequilibrium quantum dynamics has yet to be fully understood.

An ultracold Fermi gas provides a paradigm for scale invariant quantum system, with which the atoms interactions can be tunable between two scale invariant regimes, from an ideal, noninteracting gas to the most strongly interacting, nonrelativistic quantum system ever known~\cite{Son}. This conformal and scale invariance has leading to two important consequences. One is that both gases have a universal thermodynamics that the pressure $p$ and the energy density $\epsilon$ are related by $p = 2 /3 \,\epsilon$ \cite{HoUniversalThermo, ThomasUniversal}. Another remarkable consequence is that the bulk viscosity  vanishes for arbitrary temperatures in such quantum gases~\cite{viscocity1,viscocity2,viscocity3,viscocity4}.

Recently, based on the dynamical symmetry, we have found the novel dynamics and an Efimovian expansion is investigated in such scale invariant Fermi gases \cite{Wu2}. When the trap frequency $\omega(t)$ decreases as $\omega(t)=1/(\lambda t)$ (where $\lambda$ is a controllable scale factor), the  expanded cloud size as a function of time shows up the same geometric scaling behavior as the well-known three-body Efimov effect~\cite{efimov,braaten,Grimm1,Jochim,Gross,Hulet,Grimm2,Ohara,Modugno,Huston,Modugno2,Minardi,Jin, Grimm3}.

In parallel to Efimov effect for three resonantly interacting non-relativistic bosons, a profound new theory for fermions, super Efimov effect,  was developed since 2013 \cite{superEfimov}. There are also an infinite number of three-body bound states when the pairwise interaction of three identical fermions in two spatial dimensions is at the vicinity of a p-wave resonance.  Remarkably, in contrast to Efimov effect where the binding energies obey a single exponential scaling,  the bound energy of the super Efimov effect follows a fascinating double exponential scaling $E_n=E_0 e^{-2e^{\pi n/s_0+\theta_0}}$, where $s_0=4/3$ and $\theta_0$ is a parameter from the short range potential. The effect could be  attributed to a universal three-body effective potential within the hyper-spherical formalism~\cite{superYu}.
 The large scale factor and double exponential scaling make its experimental realization difficult.  Although there are some proposals to use the ultracold gases of two different species with unequal masses to reduce scale factor, the observation of super Efimov effect has remaining elusive so far.
\begin{figure}
\begin{center}\
\includegraphics[width=85mm]{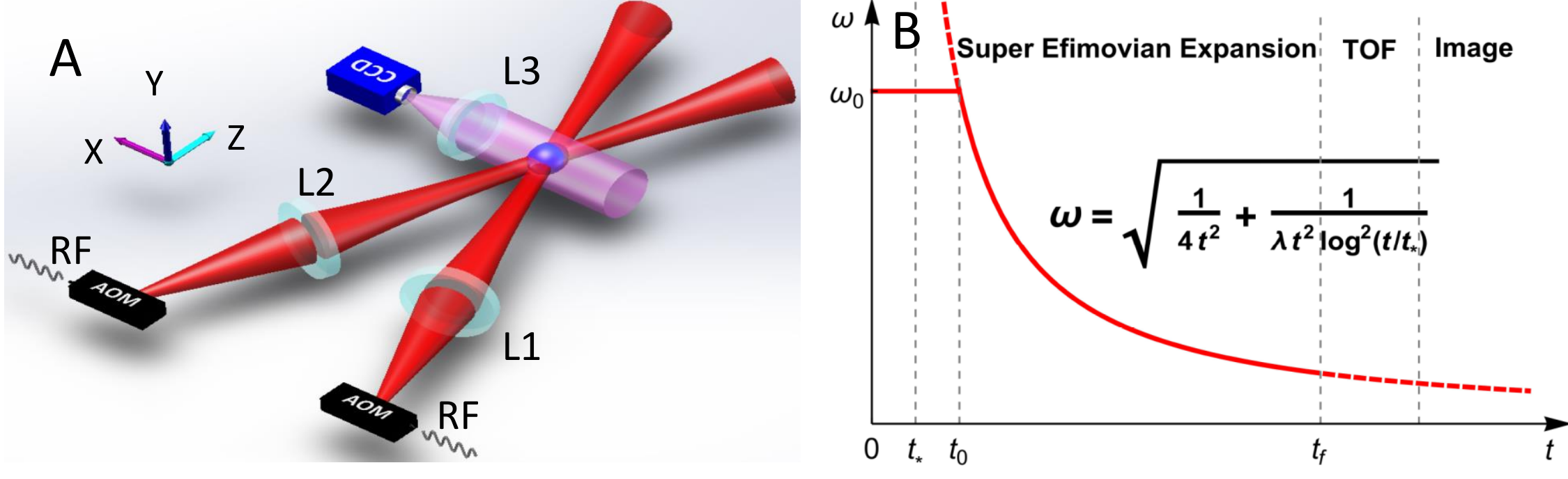}
\end{center}
\caption{The schematic of experimental setup (A) and the super Efimovian expansion (B). A scale invariant ultracold gas is held in a harmonic trap consisted of the two dipole-trap laser beams with frequency $\omega_{0}$. Then, with an initial time $t_0$, the trap frequency starts to decrease as $\omega(t)=[1/4t^2+1/t^2\lambda\log(t/t_*)]^{1/2}$, which $t_*$ is a control parameter.  L1-L3, achromatic lenses; CCD, charge coupled device; AOM; acousto-optic modulator.
\label{fig:Cloudsize}}
\end{figure}
\begin{figure*}[t]
  \begin{minipage}{.85\textwidth}
    \includegraphics[angle=0,width=\textwidth,height=3.8 in]
    {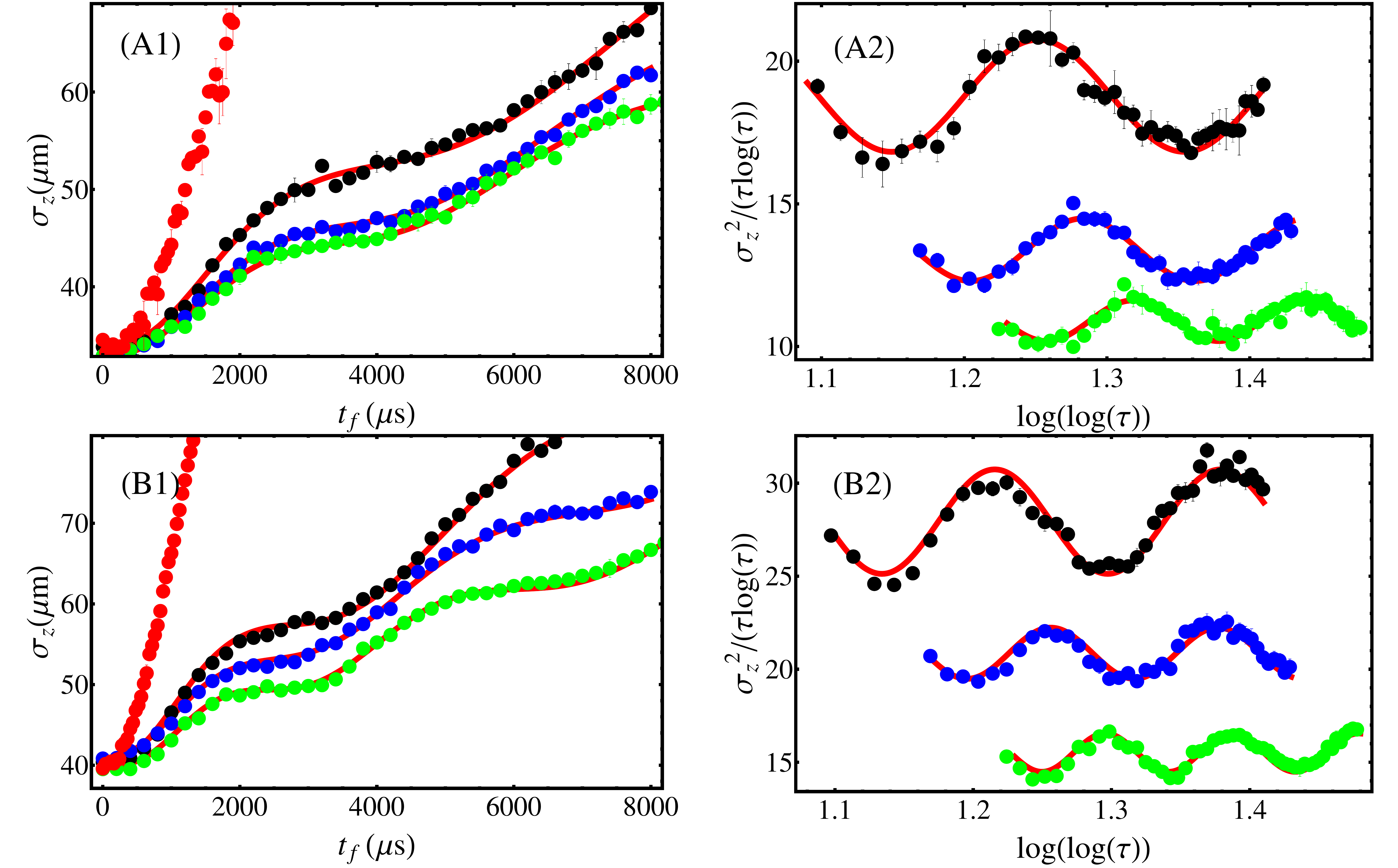}
  \end{minipage}
        \caption{\label{expansion}
   The expansion dynamics of the cloud. $A_1$ ($B_1$) and $A_2$ ($B_2$) are the mean axial cloud size $\sigma_z$  versus time and the dimensionless mean square axial cloud size $\sigma_z^2/(\tau\log(\tau))$ verse the dimensionless time $\log(\log(\tau))$ at $B=832$ Gauss; unitary Fermi gas ($B=528$ Gauss; an ideal non-interacting Fermi gas), respectively, where $\tau=(t+t_0)/t_*$. Black, blue, and green  dots are measured data with $t_0=4$ ms ($\lambda_z=2.7\times 10^{-3}$), $t_0=5$ ms ($\lambda_z=1.5\times 10^{-3}$) and $t_0=6$ ms ($\lambda_z=9\times 10^{-4}$), respectively. Red dots are the expansion size of the cloud when $\lambda_z$is chosen as the critical point $\lambda_z=4$. Solid red curves for the theoretical fits based on Eq. 1. Here $t_*=0.2$ ms. Error bars represent the standard deviation of the statistic.
}
\end{figure*}

Very recently, the super Efimovian dynamics in scale invariant Fermi gases is proposed~\cite{Super}.  By controlling the trap frequency of the external harmonic trap as $\omega(t)=[1/4t^2+1/t^2\lambda\log(t/t_*)]^{1/2}$, the double-log periodicity of the cloud expansion emerges when $\lambda<4$. The cloud dynamics is then given by
\begin{equation}
\langle R^2\rangle=A t\log(t/t_*)\{1+B\cos[s_0\log\bigg(\log(t/t_*)\bigg)+\phi]\},
\end{equation}
where $\langle R^2\rangle $ is the mean square cloud size and $t_*$ is the tunable parameters. A, B and $\phi$ are constants determined by the initial conditions. $s_0\equiv\omega_b\sqrt{1/\lambda-1/4}$ where $\omega_b$ is a factor related to the breathing mode frequency; $\omega_b = 2$ for the non-interacting gas and $\omega_b=\sqrt{12/5}$ for the anisotropic unitary Fermi gas along the axial direction. In this letter,  we report the first observation for such the super Efimovian expansion dynamics in the  unitary Fermi gas.  An essential feature of the super Efimov effect, double-log periodicity, is revealed by measuring the cloud size in the expansion.  The novel dynamics could be observed in any other scale invariant quantum gases. The universality is verified both in the non-interacting limit and in the unitarity limit. Compared to previous Efimiovian expansion~\cite{Wu2}, observing super-Efmovian evolution in such scale invariant Fermi gases requires more precisely controlling the external harmonic trap frequencies and represents a paradigm in probing universal dynamics and allows in a new way benchmarking with intriguing double-exponential scaling symmetry.

The experimental setup is similar to that Ref.~\cite{Wu1,Wu2} and the schematic is shown in Fig. 1A. We used a balanced mixture of $^6$Li fermions in the two lowest hyperfine states $|\uparrow\rangle\equiv|F=1/2, M=-1/2\rangle$ and $|\downarrow\rangle\equiv|F=1/2, M= 1/2\rangle$. The fermionic atoms are loaded into a cross-dipole trap to perform evaporative cooling~\cite{Wu1}. The resulting potential has a cylindrical symmetry with the trap anisotropic frequency ratio $\omega_r/\omega_z$  about 10. The magnetically induced collisonal resonance is used to tune the interaction of the atoms either at unitary regime with the magnetic field $B=832$ Gauss or at ideal non-interacting regime with $B=528$ Gauss. The system is initially prepared in a stationary state at unitary with the trap depth fixed at $0.5\%\,U_0$ for effectively cooling where $U_0$ is the full trap potential depth. The trap depth is then raised to $2\, \% U_0$  and held for several hundreds of milliseconds for equilibrium of the gas. The energy of Fermi gas is $E=0.8\,E_F$, corresponding to the temperature $T=0.25\,T_F$, which $E_F$ and $T_F$ are the Fermi energy and temperature of an ideal Fermi gas, respectively. The initial axial and radial trap frequency are $\omega_{z0}=2 \pi\times255.8$ Hz and $\omega_{r0}=2 \pi\times2538.4$ Hz, respectively.  Subsequently the trap frequencies are lowered with the relation as $\sqrt{1/4t^2+1/t^2\lambda\log(t/t_*)}$ to allow the cloud dynamical expansion (Fig. 1B). After a time of evolution $t_f$ in the time-dependent trap, the trap beams are completely turned off and the cloud is probed via standard resonant absorption imaging techniques after a time-of-flight (tof) expansion time $t_{\rm tof}=400\,\mu s$. The density profile along the axial(radial) direction is fitted by a Gaussian function, from which we obtain $\sigma_{z,obs}$($\sigma_{r,obs}$). $\sigma_{z,obs}$ ($\sigma_{r,obs}$) is related to the in-situ cloud size by a scale factor which can be obtained by either hydrodynamic or ballistic expansion equation with the time-of-flight time $t_{tof}$ (see the supplementary material of Ref. \cite{Wu2} for detail), with which  the mean square cloud size $\sigma_z$ ($\sigma_r$) can be obtained. Each data point is  averaged over $5$ shots taken with identical parameters to reduce statistic errors.

Figure 2 shows the typical measurements of the dependence of the axial cloud size $\sigma_z$ on the evolution time $t_f$  for both the unitary ($B=832$ Gauss) and the noninteracting Fermi gases ($B=528$ Gauss), where the control parameter $t_*$ is set as  $t_*=0.2 $ ms. In the experiment, the axial trap frequency is decreased from an stationary state. The initial trap frequency  $\omega_{z0}$ is connected with  an initial time $t_0$ and the control parameter $t_*$ as $\lambda_z=1/[(\omega_{z0}^2t_0^2-1/4)\log^2(t_0/t_*)^2]$. Black, blue and green dots are the measured data with different $t_0$ ($\lambda_z$). The solid red lines are the theoretical curves based on Eq. 1 with  $s_0$ as fitting parameter. Because the gas is in the normal fluid regime and therefore $\sigma_z$ is obtained by a Gaussian fit to the density profile, $\sigma_z = 2\sqrt{\langle R^2\rangle}$  and thus the theoretical expression for $\sigma_{z}$ is simply a square root of Eq. 1. The expansion dynamics of the cloud for $\lambda_z<4$ (black, blue and green dots) are evidently different with one for $\lambda_z\gg 4$ (red dots denote as the expansion for $\lambda_z=4$). The expansion shows the signature of the super Efimov effect.  In order to better show this novel dynamics, the dimensionless mean square axial cloud size $\sigma_z^2/(\log(\tau))$ as a function of the dimensionless time $\log(\log(\tau))$ at the unitary Fermi gas  and the ideal non-interacting Fermi gas are plotted in Fig. 2$B_1$ and Fig. 2$B_2$, respectively, where $\tau=(t_f+t_0)/t_*$. It is clearly shown that the double-log periodicity, a hallmark of the super Efimov effect,  is emergent for the cloud expansion dynamics with  smaller $\lambda_z$ and there is an excellent agreement between theory and experiment.

Note that the aspect ratio of the trap frequencies remains unchanged when the trap frequency is controlled with $\omega(t)=1/(\lambda t)$ for an Efimov expansion ~\cite{Wu2}. To perform the super Efimov expansion, the aspect ratio of the trap frequencies should be changed due to the logarithmic dependence on time of the trap frequency. In the experiment, however,  the parameters are carefully chosen so that the change of aspect ratio, $\omega_r/\omega_z$, is negligible in the expansion dynamics.
\begin{figure}[tb]
\begin{center}
\includegraphics[width=3.0 in]{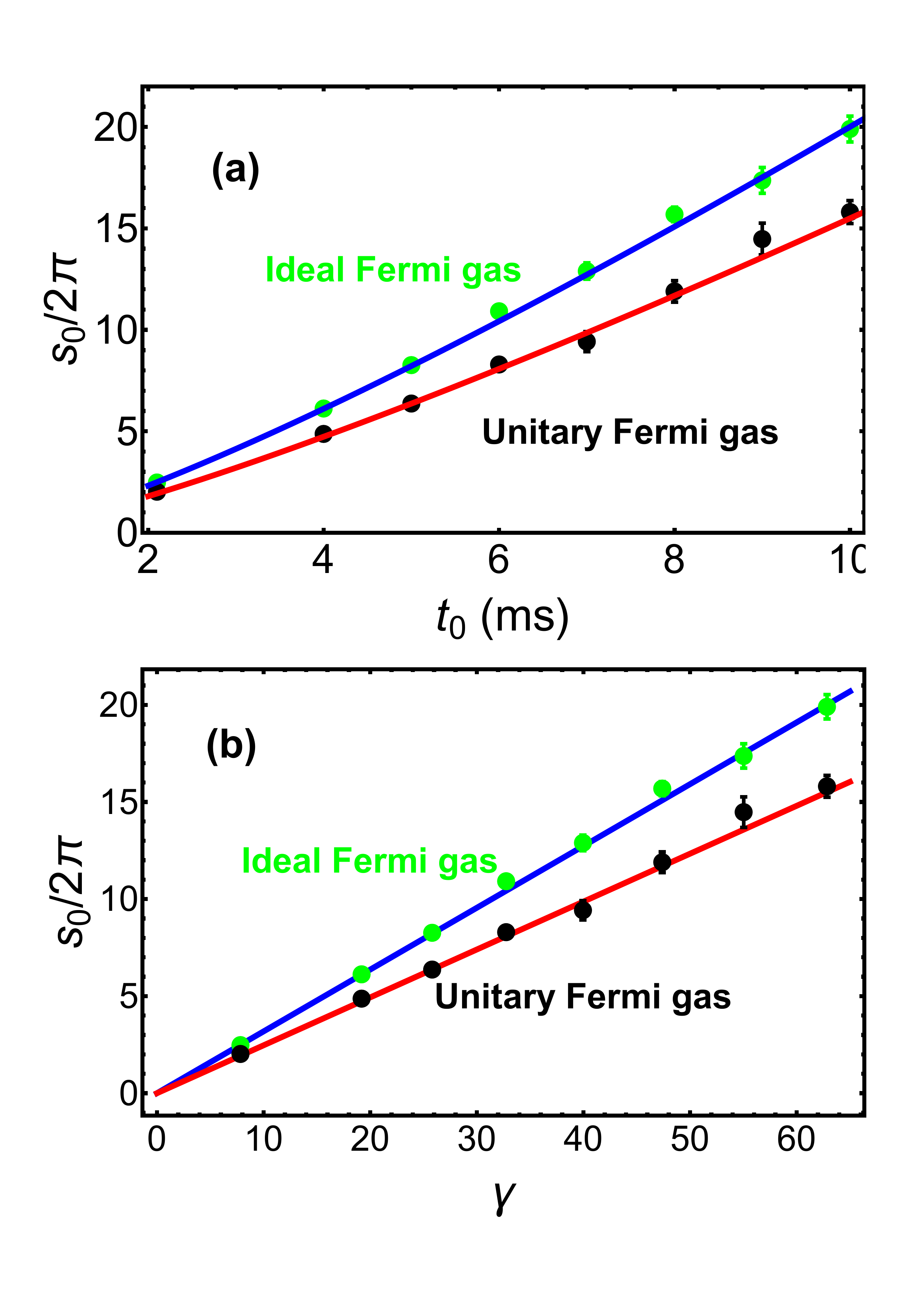}
\end{center}
\caption{The universality of the expansion. (a) $s_0$ as a function of $t_0$. (b) $s_0$ as a function of $\gamma$, where $\gamma=\sqrt{1/\lambda_z-1/4}$.  The solid lines are the linear fitting curves and the dashed lines are $s_0 = \omega_b \gamma$  with $\omega_b = 2$ for the noninteracting fermions and $\omega_b =\sqrt{12/5}$ for the unitary Fermi gas. Error bars represent the standard deviation of the statistic. The other parameters are $\omega_0=2\pi\times255.8$ Hz and $t_*=0.2$ ms.
\label{universal}}
\end{figure}

Now we demonstrate that this super Efiomovian expansion dynamics is universal. With Eq. (1), the oscillation period of the expansion dynamics is depended on the relation of $s_0$ and $\lambda_z$. Four tunable parameters $\omega_{z0}$, $t_0$, $\lambda_z$ and $t_*$ are related as $\lambda_z=1/[(\omega_{z0}^2t_0^2-1/4)\log^2(t_0/t_*)]$ in which three parameters are independent. Therefore, except that $s_0=\omega_b(1/\lambda-1/4)^{1/2}$, $s_0$ can also be determined by the parameters $t_0$, $t_*$ and $\omega_0$ as follows,
\begin{equation}
s_0=\omega_b\bigg(\omega_{z0}^2t_0^2\log^2(t_0/t_*)-1/4(1+\log^2(t_0/t_*))\bigg)^{1/2}.
\label{universal2}
\end{equation}

We demonstrate such universal relations in the experiment.  $\lambda_z$ is determined by the trap frequencies measured by the parametric resonance and $t_*$ is set as $0.2$ ms. $s_0$ is extracted from the best fit of the expansion data in Fig. \ref{expansion}. The universal relation between $s_0$ and  $t_0$ is plotted in Fig. \ref{universal}(a), where the parameters $\omega_0$ and $t_*$ are fixed ($\omega_0=2\pi\times255.8$ Hz and $t_*=0.2$ ms). The solid curves in Fig.  \ref{universal}(a) are the calculated results with Eq. \ref{universal2} without any free parameters. The experimental measurement is agreed very well with the theoretical prediction.

Fig. \ref{universal}(b) shows the universal relation between  $s_0$ and $\gamma$, where $\gamma\equiv\sqrt{1/\lambda_z-1/4}$. The green and black dots are the measurements for the ideal Fermi gas and unitray Fermi gas, respectively. The data, $s_0(\gamma)$,  can fit very well with a linear function $s_0=\kappa \gamma$ which gives the slope $\kappa=2.02\pm0.02$ for the non-interacting case and $\kappa=1.55\pm0.03$ for the unitary Fermi gas. These are in good agreement with $\omega_b=2$ for the non-interacting case and $\omega_b=\sqrt{12/5}=1.55$ for the unitary case. The such super Efimovian expansion is also robust and insensitive to the temperature and atoms number of the Fermi gas. The dependence of  $s_0$ of the unitary Fermi gas on these parameters, where  $\lambda_z$ and $t_*$ are fixed and both the temperature and the atom number are varied by controlling the evaporative cooling process. The measured results show that $s_0$ is almost a constant for  changing of the atoms number of each spin component  and the dimensionless temperature $T/T_F$.

\begin{figure}[tb]
\begin{center}
\includegraphics[width=2.6 in]{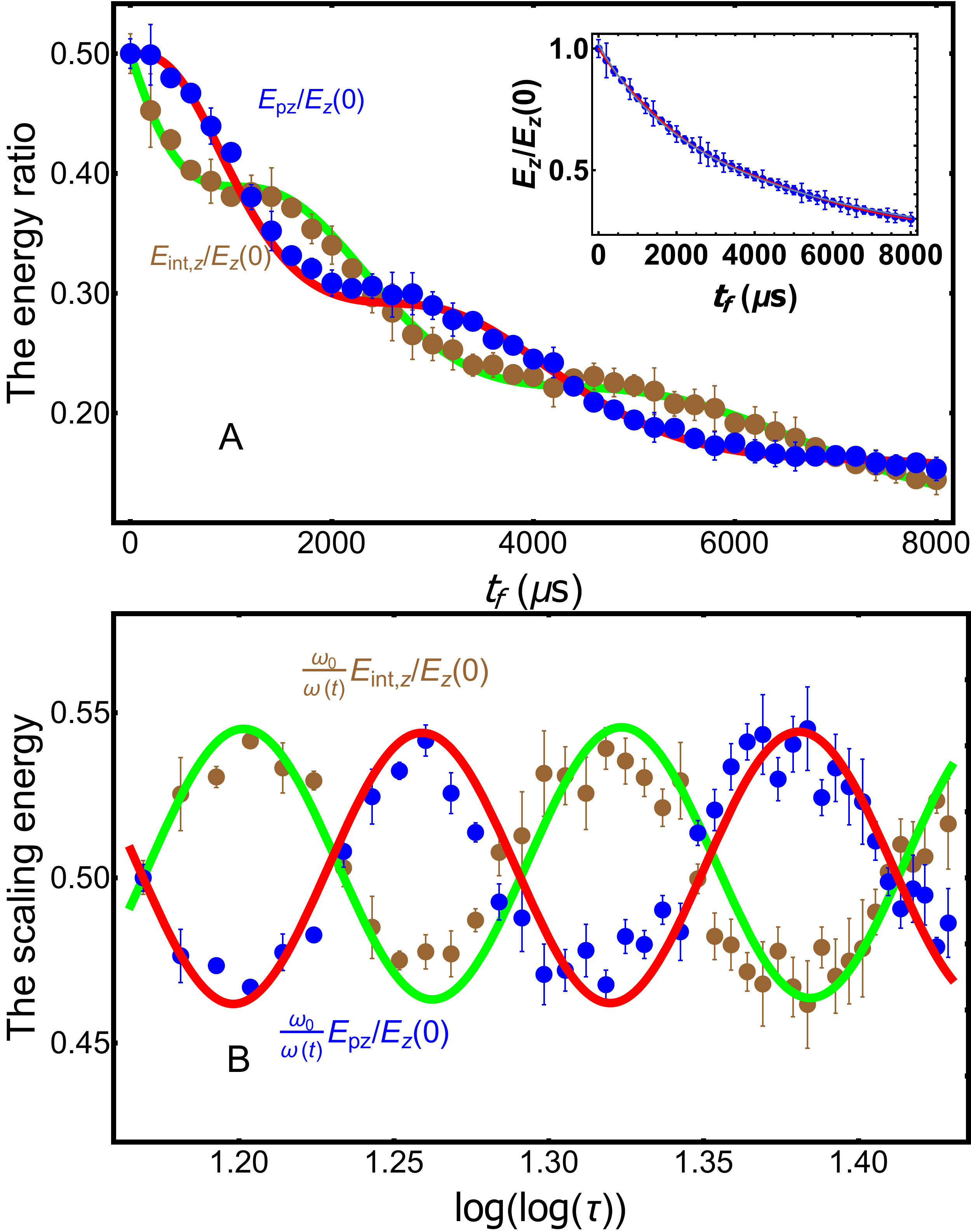}
\end{center}
\caption{The axial potential(internal) energy ratio verse $t_f$ (A) and the axial energy scaling verse $\log(\log(\tau))$ (B) for the super Efimovian expansion. The dots are measured data and the solid curves are the theoretical prediction. The blue dots are the scaled potential energy  and brown dots are the scaled internal energy. Inset is the dependence of the total energy on the expansion time.  Error bars represent the standard deviation of the statistic. The other parameters are the same as Fig. 2.
\label{energy}}
\end{figure}

Last, we investigate the energy scaling in the super Efimovian expansion. The energy  $E$ of the trapped gas is the sum of potential energy $E_p$ and internal energy $E_{int}$ (sum of kinetic and interaction energies), $E=E_p+E_{int}$. Our experimental approach to measure the energy of a two-component Fermi gas is based on the analysis of time of flight (TOF) images of atoms released from an anisotropic trap. The potential energy could be derived from the Hamiltonian and is  given by
\begin{equation}
E_p=\sum_{j=x,y,z}E_{p,j}=\frac{1}{2} \sum_{j}\frac{\omega^2_j(t)}{\omega^2_{j0}}\frac{R^2_j(t)}{R^2_j(0)}E_j(0),
\end{equation}
where $E_{p,j}$ and $E_{j}(0)$ are the potential energy and initial energy along the $j$ direction. Due to scale invariance, the  dependence of internal energy on time can be written as $d E_{int}/d t= -1/2 m\sum_{j}\omega_j(t)^2 dR_j^2(t)/d t$. Using the scaling equations~\cite{scalingequation1,scalingequation2}, the internal energy is obtained by
\begin{equation}
 E_{int}=\sum_{j=x,y,z}E_{int,j}=\frac{1}{2}\sum_{j}\bigg(\frac{R_j(0)^2}{R^2_j(t)}+\frac{\omega^2_j(t)}{\omega^2_{j0}}\frac{\dot{R}^2_j(t)}{R^2_j(0)}\bigg)E_j(0).
 \end{equation}
 Therefore, the ratio between the potential(internal) energy and the initial energy of trapped gas is determined by the mean cloud size and trap frequency. The measured axial potential and internal energy dependence on the expansion time for ideal Fermi gas is presented  by Fig. \ref{energy}A. The measured data (dots) are consistent with  the theoretical predictions (solid curves). It is clearly that the change of the potential energy and internal energy is not trivial and alternately exchanged as the expansion time. The period of the energy exchange is close to the oscillated period of the super Efimovian expansion. However, the total energy is  monotonously decreased (inset). Fig. \ref{energy}B shows the axial potential (internal) energy scaling as a function of $\log(\log(\tau))$. It is evident that both potential and internal energy are also double-log periodicity with an $\pi/2$ phase difference in the super Efimovian expansions. The experimental data are agreed well with the calculations(solid curves).

In conclusion, we study the super Efimovian dynamical expansion in strongly interacting Fermi gas by engineering the trapped frequencies.  The double-log periodicity, an essential feature for super Efimov effect, is observed and investigated. The universality of such novel dynamics expansion is domenstrated on the ideal and unitary Fermi gas. The super Efimovian dynamical expansion can be studied for all scale invariant Fermi gas. The Observing super-Efmovian evolution in such scale invariant Fermi gases provides a new way to probing universal dynamics and and the intriguing scaling symmetry.

We thank Z. Yu and  H. Zhai for helpful discussions. This research is supported by the National Key Reserach and Development Program of China (grant no. 2017YFA0304201) and National Natural Science Foundation of China (NSFC) (grant nos. 11374101 and 91536112).


\begin{thebibliography}{23}

\bibitem{ScaleThermal1}
K. G. Wilson, Rev. Mod. Phys. {\bf 55}, 583 (1983).

\bibitem{ScaleThermal2}
 S. Sachdev, Quantum Phase Transitions, (Cambridge University Press 2011).
 \bibitem{critialpoint}
P. C. Hohenberg, B. I. Halperin, Rev. Mod. Phys. {\bf 49}, 435 (1977).
\bibitem{Scalebio}
J. H. Brown, G. B.West (eds), Scaling in Biology (Oxford University Press, 2000).





\expandafter\ifx\csname natexlab\endcsname\relax\def\natexlab#1{#1}\fi
\expandafter\ifx\csname bibnamefont\endcsname\relax
  \def\bibnamefont#1{#1}\fi
\expandafter\ifx\csname bibfnamefont\endcsname\relax
  \def\bibfnamefont#1{#1}\fi
\expandafter\ifx\csname citenamefont\endcsname\relax
  \def\citenamefont#1{#1}\fi
\expandafter\ifx\csname url\endcsname\relax
  \def\url#1{\texttt{#1}}\fi
\expandafter\ifx\csname urlprefix\endcsname\relax\def\urlprefix{URL }\fi
\providecommand{\bibinfo}[2]{#2}
\providecommand{\eprint}[2][]{\url{#2}}

\bibitem[{\citenamefont{Nishida and Son}(2007)}]{Son}
\bibinfo{author}{\bibfnamefont{Y.}~\bibnamefont{Nishida}} \bibnamefont{and}
  \bibinfo{author}{\bibfnamefont{D.}~\bibnamefont{Son}},
  \bibinfo{journal}{Phys. Rev. D} \textbf{\bibinfo{volume}{76}},
  \bibinfo{pages}{086004} (\bibinfo{year}{2007}).

\bibitem[{\citenamefont{Ho}(2004)}]{HoUniversalThermo}
\bibinfo{author}{\bibfnamefont{T.-L.} \bibnamefont{Ho}},
  \bibinfo{journal}{Phys. Rev. Lett.} \textbf{\bibinfo{volume}{92}},
  \bibinfo{pages}{090402} (\bibinfo{year}{2004}).

\bibitem[{\citenamefont{Thomas et~al.}(2005)\citenamefont{Thomas, Kinast, and
  Turlapov}}]{ThomasUniversal}
\bibinfo{author}{\bibfnamefont{J.~E.} \bibnamefont{Thomas}},
  \bibinfo{author}{\bibfnamefont{J.}~\bibnamefont{Kinast}}, \bibnamefont{and}
  \bibinfo{author}{\bibfnamefont{A.}~\bibnamefont{Turlapov}},
  \bibinfo{journal}{Phys. Rev. Lett.} \textbf{\bibinfo{volume}{95}},
  \bibinfo{pages}{120402} (\bibinfo{year}{2005}).

  \bibitem{viscocity1}
D. T. Son, Phys. Rev. Lett. {\bf 98}, 020604 (2007)
\bibitem{viscocity2}
M. A. Escobedo, M. Mannarelli, and C. Manuel, Phys. Rev. A {\bf 79}, 063623 (2009).

\bibitem{viscocity3}
Y. H. Hou, L. P. Pitaevskii, and S. Stringari, Phys. Rev. A {\bf 87}, 033620 (2013).

\bibitem{viscocity4}
K. Dusling and T. Schafer, Phys. Rev. Lett. {\bf 111}, 120603 (2013).

\bibitem[{\citenamefont{Deng et~al.}(2016)\citenamefont{Deng, Diao, Yu, and
  Wu}}]{Wu2}
\bibinfo{author}{\bibfnamefont{S.}~\bibnamefont{Deng}},
 \bibinfo{author}{\bibfnamefont{Z.-Y}~\bibnamefont{Shi}},
  \bibinfo{author}{\bibfnamefont{P.}~\bibnamefont{Diao}},
  \bibinfo{author}{\bibfnamefont{Q.}~\bibnamefont{Yu}},
   \bibinfo{author}{\bibfnamefont{H.}~\bibnamefont{Zhai}},
    \bibinfo{author}{\bibfnamefont{R.}~\bibnamefont{Qi}},~\bibnamefont{and}
  \bibinfo{author}{\bibfnamefont{H.}~\bibnamefont{Wu}}, \bibinfo{journal}{Science} \textbf{\bibinfo{volume}{371}}, \bibinfo{pages}{374}
  (\bibinfo{year}{2016}).

\bibitem[{\citenamefont{Efimov}(1970)}]{efimov}
\bibinfo{author}{\bibfnamefont{V.}~\bibnamefont{Efimov}},
  \bibinfo{journal}{Phys. Lett.} \textbf{\bibinfo{volume}{33B}},
  \bibinfo{pages}{563} (\bibinfo{year}{1970}).
\bibitem[{\citenamefont{Braaten and Hammer}(2006)}]{braaten}
\bibinfo{author}{\bibfnamefont{E.}~\bibnamefont{Braaten}} \bibnamefont{and}
  \bibinfo{author}{\bibfnamefont{H.}~\bibnamefont{Hammer}},
  \bibinfo{journal}{Phys.Rep.} \textbf{\bibinfo{volume}{428}},
  \bibinfo{pages}{259} (\bibinfo{year}{2006}).


\bibitem[{\citenamefont{Kraemer et~al.}(2006)\citenamefont{Kraemer, Mark,
  Waldburger, Danzl, Chin, Engeser, Lange, Pilch, Jaakkola, Nagerl
  et~al.}}]{Grimm1}
\bibinfo{author}{\bibfnamefont{T.}~\bibnamefont{Kraemer}},
  \bibinfo{author}{\bibfnamefont{M.}~\bibnamefont{Mark}},
  \bibinfo{author}{\bibfnamefont{P.}~\bibnamefont{Waldburger}},
  \bibinfo{author}{\bibfnamefont{J.}~\bibnamefont{Danzl}},
  \bibinfo{author}{\bibfnamefont{C.}~\bibnamefont{Chin}},
  \bibinfo{author}{\bibfnamefont{B.}~\bibnamefont{Engeser}},
  \bibinfo{author}{\bibfnamefont{A.}~\bibnamefont{Lange}},
  \bibinfo{author}{\bibfnamefont{K.}~\bibnamefont{Pilch}},
  \bibinfo{author}{\bibfnamefont{A.}~\bibnamefont{Jaakkola}},
  \bibinfo{author}{\bibfnamefont{H.}~\bibnamefont{Nagerl}},
  \bibnamefont{and}
  \bibinfo{author}{\bibfnamefont{R.}~\bibnamefont{Grimm}},
  \bibinfo{journal}{Nature}
  \textbf{\bibinfo{volume}{440}}, \bibinfo{pages}{315} (\bibinfo{year}{2006}).

\bibitem[{\citenamefont{Ottenstein et~al.}(2008)\citenamefont{Ottenstein,
  Lompe, Kohnen, Wenz, and Jochim}}]{Jochim}
\bibinfo{author}{\bibfnamefont{T.}~\bibnamefont{Ottenstein}},
  \bibinfo{author}{\bibfnamefont{T.}~\bibnamefont{Lompe}},
  \bibinfo{author}{\bibfnamefont{M.}~\bibnamefont{Kohnen}},
  \bibinfo{author}{\bibfnamefont{A.}~\bibnamefont{Wenz}}, \bibnamefont{and}
  \bibinfo{author}{\bibfnamefont{S.}~\bibnamefont{Jochim}},
  \bibinfo{journal}{Phys. Rev. Lett.} \textbf{\bibinfo{volume}{101}},
  \bibinfo{pages}{203202} (\bibinfo{year}{2008}).

\bibitem[{\citenamefont{Gross et~al.}(2009)\citenamefont{Gross, shotan,
  Kokkelmans, and Khaykovich}}]{Gross}
\bibinfo{author}{\bibfnamefont{N.}~\bibnamefont{Gross}},
  \bibinfo{author}{\bibfnamefont{Z.}~\bibnamefont{Shotan}},
  \bibinfo{author}{\bibfnamefont{S.}~\bibnamefont{Kokkelmans}},
  \bibnamefont{and}
  \bibinfo{author}{\bibfnamefont{L.}~\bibnamefont{Khaykovich}},
  \bibinfo{journal}{Phys. Rev. Lett.} \textbf{\bibinfo{volume}{103}},
  \bibinfo{pages}{163202} (\bibinfo{year}{2009}).

\bibitem[{\citenamefont{Pollack et~al.}(2009)\citenamefont{Pollack, Dries, and
  Hulet}}]{Hulet}
\bibinfo{author}{\bibfnamefont{S.}~\bibnamefont{Pollack}},
  \bibinfo{author}{\bibfnamefont{D.}~\bibnamefont{Dries}}, \bibnamefont{and}
  \bibinfo{author}{\bibfnamefont{R.}~\bibnamefont{Hulet}},
  \bibinfo{journal}{Science} \textbf{\bibinfo{volume}{326}},
  \bibinfo{pages}{1683} (\bibinfo{year}{2009}).

\bibitem[{\citenamefont{Knoop et~al.}(2009)\citenamefont{Knoop, Ferlaino, Mark,
  Berninger, Schobel, Nagerl, and Grimm}}]{Grimm2}
\bibinfo{author}{\bibfnamefont{S.}~\bibnamefont{Knoop}},
  \bibinfo{author}{\bibfnamefont{F.}~\bibnamefont{Ferlaino}},
  \bibinfo{author}{\bibfnamefont{M.}~\bibnamefont{Mark}},
  \bibinfo{author}{\bibfnamefont{M.}~\bibnamefont{Berninger}},
  \bibinfo{author}{\bibfnamefont{H.}~\bibnamefont{Schobel}},
  \bibinfo{author}{\bibfnamefont{H.}~\bibnamefont{Nagerl}}, \bibnamefont{and}
  \bibinfo{author}{\bibfnamefont{R.}~\bibnamefont{Grimm}},
  \bibinfo{journal}{Nat. Phys.} \textbf{\bibinfo{volume}{5}},
  \bibinfo{pages}{227} (\bibinfo{year}{2009}).

\bibitem[{\citenamefont{Williams et~al.}(2009)\citenamefont{Williams, Hazlett,
  Huckans, Stites, Zhang, and O'Hara}}]{Ohara}
\bibinfo{author}{\bibfnamefont{J.}~\bibnamefont{Williams}},
  \bibinfo{author}{\bibfnamefont{E.}~\bibnamefont{Hazlett}},
  \bibinfo{author}{\bibfnamefont{J.}~\bibnamefont{Huckans}},
  \bibinfo{author}{\bibfnamefont{R.}~\bibnamefont{Stites}},
  \bibinfo{author}{\bibfnamefont{Y.}~\bibnamefont{Zhang}}, \bibnamefont{and}
  \bibinfo{author}{\bibfnamefont{K.}~\bibnamefont{O'Hara}},
  \bibinfo{journal}{Phys. Rev. Lett.} \textbf{\bibinfo{volume}{103}},
  \bibinfo{pages}{130404} (\bibinfo{year}{2009}).

\bibitem[{\citenamefont{Zaccanti et~al.}(2009)\citenamefont{Zaccanti, Deissler,
  D'Errico, Fattori, Jona-Lasinio, Muller, Roati, Inguscio, and
  Modugno}}]{Modugno}
\bibinfo{author}{\bibfnamefont{M.}~\bibnamefont{Zaccanti}},
  \bibinfo{author}{\bibfnamefont{B.}~\bibnamefont{Deissler}},
  \bibinfo{author}{\bibfnamefont{C.}~\bibnamefont{D'Errico}},
  \bibinfo{author}{\bibfnamefont{M.}~\bibnamefont{Fattori}},
  \bibinfo{author}{\bibfnamefont{M.}~\bibnamefont{Jona-Lasinio}},
  \bibinfo{author}{\bibfnamefont{S.}~\bibnamefont{Muller}},
  \bibinfo{author}{\bibfnamefont{G.}~\bibnamefont{Roati}},
  \bibinfo{author}{\bibfnamefont{M.}~\bibnamefont{Inguscio}}, \bibnamefont{and}
  \bibinfo{author}{\bibfnamefont{G.}~\bibnamefont{Modugno}},
  \bibinfo{journal}{Nat. Phys.} \textbf{\bibinfo{volume}{5}},
  \bibinfo{pages}{586} (\bibinfo{year}{2009}).

\bibitem[{\citenamefont{Berninger et~al.}(2011)\citenamefont{Berninger,
  Zenesini, Huang, Harm, Nagerl, Ferlaino, Grimm, Julienne, and
  Huston}}]{Huston}
\bibinfo{author}{\bibfnamefont{M.}~\bibnamefont{Berninger}},
  \bibinfo{author}{\bibfnamefont{A.}~\bibnamefont{Zenesini}},
  \bibinfo{author}{\bibfnamefont{B.}~\bibnamefont{Huang}},
  \bibinfo{author}{\bibfnamefont{W.}~\bibnamefont{Harm}},
  \bibinfo{author}{\bibfnamefont{H.}~\bibnamefont{Nagerl}},
  \bibinfo{author}{\bibfnamefont{F.}~\bibnamefont{Ferlaino}},
  \bibinfo{author}{\bibfnamefont{R.}~\bibnamefont{Grimm}},
  \bibinfo{author}{\bibfnamefont{P.}~\bibnamefont{Julienne}}, \bibnamefont{and}
  \bibinfo{author}{\bibfnamefont{J.}~\bibnamefont{Huston}},
  \bibinfo{journal}{Phys. Rev. Lett.} \textbf{\bibinfo{volume}{107}},
  \bibinfo{pages}{120401} (\bibinfo{year}{2011}).

\bibitem[{\citenamefont{Roy et~al.}(2012)\citenamefont{Roy, Landini,
  Trenkwalder, Semeghini, Spagnolli, Simoni, Fattori, Inguscio, and
  Modugno}}]{Modugno2}
\bibinfo{author}{\bibfnamefont{S.}~\bibnamefont{Roy}},
  \bibinfo{author}{\bibfnamefont{M.}~\bibnamefont{Landini}},
  \bibinfo{author}{\bibfnamefont{A.}~\bibnamefont{Trenkwalder}},
  \bibinfo{author}{\bibfnamefont{G.}~\bibnamefont{Semeghini}},
  \bibinfo{author}{\bibfnamefont{G.}~\bibnamefont{Spagnolli}},
  \bibinfo{author}{\bibfnamefont{A.}~\bibnamefont{Simoni}},
  \bibinfo{author}{\bibfnamefont{M.}~\bibnamefont{Fattori}},
  \bibinfo{author}{\bibfnamefont{M.}~\bibnamefont{Inguscio}}, \bibnamefont{and}
  \bibinfo{author}{\bibfnamefont{G.}~\bibnamefont{Modugno}},
  \bibinfo{journal}{Phys. Rev. Lett.} \textbf{\bibinfo{volume}{111}},
  \bibinfo{pages}{053202} (\bibinfo{year}{2012}).

\bibitem[{\citenamefont{Barontini et~al.}(2009)\citenamefont{Barontini, Weber,
  Rabatti, Catani, Thalhammer, Inguscio, and Minardi}}]{Minardi}
\bibinfo{author}{\bibfnamefont{G.}~\bibnamefont{Barontini}},
  \bibinfo{author}{\bibfnamefont{C.}~\bibnamefont{Weber}},
  \bibinfo{author}{\bibfnamefont{F.}~\bibnamefont{Rabatti}},
  \bibinfo{author}{\bibfnamefont{J.}~\bibnamefont{Catani}},
  \bibinfo{author}{\bibfnamefont{G.}~\bibnamefont{Thalhammer}},
  \bibinfo{author}{\bibfnamefont{M.}~\bibnamefont{Inguscio}}, \bibnamefont{and}
  \bibinfo{author}{\bibfnamefont{F.}~\bibnamefont{Minardi}},
  \bibinfo{journal}{Phys. Rev. Lett.} \textbf{\bibinfo{volume}{111}},
  \bibinfo{pages}{043201} (\bibinfo{year}{2009}).

\bibitem[{\citenamefont{Bloom et~al.}(2013)\citenamefont{Bloom, Hu, Cumby, and
  Jin}}]{Jin}
\bibinfo{author}{\bibfnamefont{R.}~\bibnamefont{Bloom}},
  \bibinfo{author}{\bibfnamefont{M.}~\bibnamefont{Hu}},
  \bibinfo{author}{\bibfnamefont{T.}~\bibnamefont{Cumby}}, \bibnamefont{and}
  \bibinfo{author}{\bibfnamefont{D.}~\bibnamefont{Jin}},
  \bibinfo{journal}{Phys. Rev. Lett.} \textbf{\bibinfo{volume}{111}},
  \bibinfo{pages}{105301} (\bibinfo{year}{2013}).

\bibitem[{\citenamefont{Huang et~al.}(2014)\citenamefont{Huang, Sidorenkov,
  Grimm, and Huston}}]{Grimm3}
\bibinfo{author}{\bibfnamefont{B.}~\bibnamefont{Huang}},
  \bibinfo{author}{\bibfnamefont{L.}~\bibnamefont{Sidorenkov}},
  \bibinfo{author}{\bibfnamefont{R.}~\bibnamefont{Grimm}}, \bibnamefont{and}
  \bibinfo{author}{\bibfnamefont{J.}~\bibnamefont{Huston}},
  \bibinfo{journal}{Phys. Rev. Lett.} \textbf{\bibinfo{volume}{112}},
  \bibinfo{pages}{190401} (\bibinfo{year}{2014}).













\bibitem{superEfimov}
Y. Nishida, S. Moroz and D. T. Son, Phys. Rev. Lett. {\bf 110}, 235301 (2013).

\bibitem{superYu}
C. Gao, J. Wang, and Z. Yu, Phys. Rev. A {\bf 92}, 020504 (2015).

 \bibitem{Super}
Z. Shi, R. Qi, H. Zhai, and Z. Yu, arXiv:1608. 05799

\bibitem[{\citenamefont{Deng et~al.}(2015)\citenamefont{Deng, Diao, Yu, and
  Wu}}]{Wu1}
\bibinfo{author}{\bibfnamefont{S.}~\bibnamefont{Deng}},
  \bibinfo{author}{\bibfnamefont{P.}~\bibnamefont{Diao}},
  \bibinfo{author}{\bibfnamefont{Q.}~\bibnamefont{Yu}}, \bibnamefont{and}
  \bibinfo{author}{\bibfnamefont{H.}~\bibnamefont{Wu}}, \bibinfo{journal}{Chin.
  Phys. Lett.} \textbf{\bibinfo{volume}{32}}, \bibinfo{pages}{053401}
  (\bibinfo{year}{2015}).

\bibitem{scalingequation1}
C. Menotti, P. Pedri, and S. Stringari, Phys. Rev. Lett. {\bf 89}, 250402 (2002).
\bibitem{scalingequation2}
K. M. O'Hara, S. L. Hemmer, M. E. Gehm, S. R. Granade, J. E. Thomas, Science {\bf 298}, 2179 (2002).




\end{thebibliography}
\end{document}